\magnification 1200\hsize 6.3 truein\vsize 8.9 truein
\def\spazio#1{\hfill\break\vbox to #1 truecm{\vfill\hbox{\hfill}}}

\def\C{\kern-10pt}
\newcount\aiuto
\newcount\pagrel
\advance\count0 by -1
\nopagenumbers\footline{
\aiuto=\count0
\advance\aiuto by \pagrel
\ifnum\aiuto = 0 \hfill
\else
{\hfill\folio\hfill}
\fi}

\vskip 3 true cm
\centerline{\bf Numerical results for generalized RVB wavefunctions.}
\centerline{\bf Application to the Hubbard model}
\vskip 3 true cm
\centerline{ G. Fano, F. Ortolani and L. Ziosi}
\vskip 0.7 true cm
\centerline{\sl  Dipartimento di Fisica, Universit\'a di Bologna}
\centerline{\sl  Via Irnerio 46 , 40126 Bologna -  Italy}
\centerline{\sl  fano@bologna.infn.it}
\vskip 0.7 true cm
\vskip 1 true cm
\vskip 0.5 true cm
%
%
%
\magnification 1200\hsize 6.3 truein\vsize 8.9 truein
\def\spazio#1{\hfill\break\vbox to #1 truecm{\vfill\hbox{\hfill}}}

\def\k{...}
  \centerline{\bf Abstract :}\par
         Numerical results are presented for a generalized resonance
valence bond state which includes both ionic and covalent contributions
in each bond ; non nearest-neighbor bonds are also considered.
 Variational calculations 
have been performed and the  space group symmetry has been
taken into account. The results for the
Hubbard Hamiltonian are compared with the exact ones for the 
$ 2 \times 2 \times 2 $  and the $ 4 \times 4 $ lattices. The agreement
is quite satisfactory for large values of the interaction U  ( for U=40
the overlap with the exact ground state wave function is 0.999 ).
Several correlation functions are also compared.\par

\noindent
\vfill\eject
\pageno=1
\noindent
{\bf 1.Introduction.}\par
\spazio 1
\par
         It is still under debate if the 2D one-band Hubbard model can
provide the basic physics of  high temperature superconductivity.
Independently of this, and in spite of the obvious oversimplification
introduced by the model in describing a strongly correlated electron
system, the theoretical challenge offered by the 2D Hubbard model is
certainly full of fascinating and unsolved problems.  Only rough
and tentative phase diagrams concerning antiferromagnetism, ferromagnetism
and superconductivity have been produced, 
manly in the large $U$  case, where the Hubbard
model maps into the $t-J$ model.\par
         It is interesting to compare the situation of our present
knowledge of the Hubbard model and the theory of the fractional
quantum Hall effect  (FQHE). Both theories refer to strongly interacting
two-dimensional electron systems;  however the situation is much more
satisfactory in the FQHE case, not only because of the simplicity
of the experimental results, but for another fundamental reason: in
the FQHE  case we have at our disposal a valid theoretical tool, 
provided by the Laughlin's wave function [1]. All subsequent theories,
like Jain's theory [2], make use of this beautiful form of the 
wave function.  \par  
         In the case of the Hubbard model, we do not have a form
of the wave function quite as satisfactory. \par
         Many years ago, Gutzwiller [3] proposed
a wave function obtained by applying a suitable operator to
the Fermi sea wave function. The physical idea is simple: one tries
to reproduce the main effect of the interaction, by weighting with
a factor  $ \eta^D $  all the configurations containing $D$ doubly
occupied sites; $\eta$ is a variational parameter.  The interesting
$ \eta = 0$  case corresponds to a state with no doubly occupied
sites  (Gutzwiller projection). \par
        The Gutzwiller wavefunction $ \psi_G $ provides a noticeable
insight into the physics of the Hubbard model. However, the 
expectation values with respect to $ \psi_G $  cannot be computed
without making some approximations. Furthermore the value of the
ground state energy, tested in finite size cases, is not more
accurate than the one given by the Hartree-Fock
 solution (spin density wave). \par
        In 1987 Anderson [4] proposed that the electronic wave
function of a high $T_c$ superconductor is a "resonating
valence-bond" (RVB) state.   This state is obtained as the
Gutzwiller projection of a BCS state with definite particle number.
It can also be considered as a superposition of valence bond states,
i.e. states describing pairs of electrons in different sites
coupled to a spin singlet.   A "covering" of the lattice is a
state in which all the electrons are coupled in pairs. \par
       A number of calculations (see, e.g. [5] ) have been performed
on the RVB state, especially in a simplified version in which
only singlets formed by nearest-neighbor electrons are allowed
(NNRVB). Almost all these works refer to the Heisenberg
and t-j two-dimensional models. Calculations involving NNRVB
states are made possible by simple graphic rules [6],[7] . \par
      In this type of approach the Gutzwiller projection is not
necessary, since the coverings of a lattice at half-filling
are products of dimers created by operators like
$ C_{ij}^+ = c_{i \uparrow}^+ c_{j \downarrow}^+ +
 c_{j \uparrow}^+ c_{i \downarrow}^+ $ ,where $i$ and $j$ are
nearest-neighbor sites and $c_{i \uparrow}^+, c_{i \downarrow}^+$
denote the usual creation operators;
 therefore no doubly occupied site occurs. \par
       It is interesting to notice that long before Anderson's
proposal of the RVB state for strongly interacting electrons, 
some theoretical chemists, Choi and Thorson [8] computed the 
one and two-particle density matrices for a generalized RVB state,
both for a linear chain and a $4 \times 4$ lattice.  By " generalized
RVB state " we mean an RVB state such that the operator  $ O^+ $
creating the valence bond between the sites $i$ and $j$  contains
not only a covalent term $ C_{ij}^+$, but also a ionic
contribution $ I_{ij}^+ =  c_{i \uparrow}^+ c_{i \downarrow}^+ + 
c_{j \uparrow}^+ c_{j \downarrow}^+ $; therefore 
$ O^+ = \lambda  C_{ij}^+ + (1-\lambda) I_{ij}^+ $, where
$ \lambda$ is a variational parameter.  These authors mentioned
the Mott insulator and the Hubbard Hamiltonian as possible 
applications, but no explicit calculations of the ground state
energy was presented.  \par
        This generalized RVB state has been proposed again for
the Hubbard model, and independently, by Messager and Richard [9];
it was also studied for a strip of $ 2 \times N $ sites by Cicuta
and Stramaglia [10].  To the best of our knowledge, there has been
no further study on this proposal, despite the following arguments
in favor of its consideration as a good variational state for the
Hubbard model: \par
1)  The state $ O^+ |0> $ is the exact ground state for a 
two electron system. \par
2)  The state $ ( O_{12}^+ O_{34}^+ - O_{13}^+ O_{24}^+ ) |0> $
is an extremely accurate approximation of the exact ground state
of a $2 \times 2$ Hubbard model. ( We recall, that in the analogous
case of the FQHE, the analytic part of the Laughlin's wave function
is a product of factors  $ (z_i - z_j)^m $, which correspond to
the {\it exact } ground state for two particles; therefore the idea
of building a many-particle wave function by combining some kind
of product of two-particle wave functions deserves particular attention.
Notice that the three particle Laughlin's w.f. is exact on the
sphere geometry.)\par
3)  The introduction of a parameter $ \lambda $ that controls
the amount of double occupancy of the bond seems to be a natural
way of constructing a wave function which covers both the large
$U$ case  ( $\lambda \to 1$ ) and the small or medium $U$ case.\par
4)  The ground state energy of the large $U$ limit of the Hubbard
model at half filling ( i.e. the Heisenberg model ) can be 
reproduced with great accuracy by including in the RVB state also
bonds connecting sites which are far apart. Indeed Liang, Doucot and 
Anderson [11] have studied this state on a bipartite lattice, by
allowing bonds connecting only sites $i$ of the sublattice $A$
with sites $j$ of the sublattice $B$, and giving a weight factor
$ h(|i - j|) $ to each bond, where the factor $h$ decreases
with a power law
with the distance $|i-j|$; notice that the positivity of $h$
guarantees the important Marshall sign rule [12] .\par
5) Even if we do not succed in finding a form of the wave
function as useful as Laughlin's w.f., we believe that a good
variational w.f. can be of use as an initial guess for modern
numerical methods. \par
     Numerical calculations on the Hubbard model can be performed
either with great accuracy on small systems, or using Montecarlo
or other approximations on larger systems. Here we will be
concerned with exact results on small systems ($2 \times 2 \times 2$
 and $4 \times 4$ ). Full
configuration interaction calculations at half filling on a
$4 \times 4$  lattice are greatly simplified by the consideration of the
space symmetry group of the model.  The best way of taking into
account the space symmetry group is to replace the electron
configurations by linear combinations of configurations contained
in a group orbit ( we recall that a group orbit is the set
$\{ g |c>, g\in G \} $ where $ |c>$ denotes a configuration ) ;
the coefficients of the linear combination are determined by the
irreducible representations of the group, and are constant
for the totally symmetric state.  The orbits are of variable 
lenght; for instance the N\'eel configuration  $ |n> $  gives 
rise to a small orbit, constituted by $|n>$ itself and the
"anti-N\'eel" configuration $ | \overline n > $ with reversed spins 
( indeed all rotations and translations of the lattice either
send  $ |n> \longrightarrow |n> $ or $ |n> \longrightarrow
 | \overline n > $ ). The dimension of the Hilbert space can be
greatly reduced since the order $ |G| $ of the space group is
large . For a square $n \times n$ lattice with periodic boundary
conditions the translation group in the
two directions has $n^2$ elements, reflections in the direction of the main
diagonals and one of the axis produce a further factor of 8; this
gives $ |G| =128 $ for n=4  ( Actually the  $4 \times 4$ lattice is
topologically equivalent to an ipercube in 4 dimensions, so a
further symmetry arises and a larger group $ \hat G $ can be found
such that  $ | \hat G | =384 $ ). For a group-theretical analysis
of the Hubbard model see ref.[13].
\par
       Besides space symmetry, the Hubbard Hamiltonian presents
the usual spin symmetry. It is a very difficult task to use an
orthogonal basis adapted to both symmetries.  One of the advantages
of the RVB approach, is that both space and spin symmetries are
introduced in the wave function from the outset. Unfortunately,
like in all valence bond methods, the wave functions to be used are
not orthogonal. \par
        In the present work we have considered
 $2 \times 2 \times 2 $ and $4 \times 4 $ systems
with periodic boundary conditions and we have computed the expectation
value of the energy and other observables (spin-spin and pairing
correlation functions ) for a generalized RVB state.  Since on a
bipartite lattice the next type of bond to be considered beside
the nearest-neighbor case is provided by bonds between electrons  at
distance $\sqrt 5 $ on the lattice ( named KM or
chess knight's move bond [14] ),
we have taken into considerations wave functions admitting also one
or more KM bonds. \par
       All the numerical results presented here correspond to 
half-filling.  However, the formalism can be easily extended  in
order to cover differnt fillings (see the end of Sec. 2 ) .
\spazio 1
{\bf 2. Formalism }.  Let us consider an $ M \times N $ square lattice.
 A site $i$ is determined by two integers $ (r,s)$. The
sublattice  $ A  $ is formed by the sites such that
$ (-1)^{r+s} = 1  $, and the sublattice $ B $ is formed by the sites
such thtat $ (-1)^{r+s} = - 1 $.  Only bonds connecting the
 sublattices  $A$ and $B$ will be allowed. \par
         We define now the creation operator of a generalized valence
bond between the sites $i$ and  $j$: 
      $$ A^+_{ij}(x,y) = x I^+_{ij} + y C^+_{ij}   \leqno (1) $$
where $x$ and $y$ are variational parameters and  $I^+_{ij}$,$C^+_{ij}$
are defined by:  
     $$  I^+_{ij} = c^+_{i\uparrow} c^+_{i\downarrow}
                   +   c^+_{j\uparrow} c^+_{j\downarrow} \leqno (2) $$
     $$  C^+_{ij} =  c^+_{i\uparrow} c^+_{j\downarrow}
                   +   c^+_{j\uparrow} c^+_{i\downarrow}. \leqno (3) $$
\par
       The operator $ C^+_{ij} $ creates the usual covalent bond, while
$ I^+_{ij} $ creates an ionic bond. Normalization of the state vector
$ A^+_{ij} (x,y) |0> $ implies the condition:
       $$  2 x^2 + 2 y^2 = 1    \leqno (4) $$
\par
        A covering of the lattice is obtained by considering the
following product of operators applied to the vacuum:
       $$ |\phi_k>  = A^+_{i_1 j_1} (x_1,y_1) A^+_{i_2 j_2} (x_2,y_2)
                 \dots A^+_{i_n j_n} (x_n,y_n) \ |0> \leqno (5) $$   
where $i_1,i_2, \dots i_n$ belong to the sublattice $A$ ,
$j_1,j_2, \dots j_n$ belong to the sublattice $B$, and the index $k$
runs over all possible coverings. \par
        At half-filling the number of lattice sites is twice the 
number of dimers, i.e.  $NM=2n$.  Because of condition (4), 
the states $|\phi_k> $ are normalized. However they are not orthogonal,
and it is also possible that they are  linearly dependent.\par
        We assume that in each factor  $ A^+_{ij} (x,y) $ the
variational parameter $x$ is a function of the distance $ |i - j| $.\par
        A generalized RVB state is a linear combination of coverings:
       $$  |\psi> \ = \ \sum_k  c_k |\phi_k> \leqno (6) $$
        Let $g$ denote an element of the space group. We assume that
the coefficients  $c_k$ are the same for all the coverings
$ g |\phi_k> , g \in G $. In this way $|\psi>$ becomes a totally
symmetric state. It must be noticed that usually the ground state
of a finite size Hubbard model is totally symmetric with respect
to the space group; an exception is provided by the
$2 \times 2$ Hubbard model. \par
       In order to compute the matrix elements of the overlap
matrix  $ S_{ij} = <\phi_i|\phi_j> $, Sutherland's graphical
rules [6] can be used. The scalar product $ <\phi_i|\phi_j> $ can
be written as $ <\phi_i|\phi_j> = \prod_{k=1}^m  l_k $, where
$l_k$ denotes the contribution of a loop with an even number of
bonds. In order to compute the factors $l_k$, the following simple
algebraic rules can be employed. From the commutator: 
    $$ [ A_{ij}(x_2,y_2), A^+_{ij}(x_1,y_1) ] = 
 2 ( \overline x_2 x_1 + \overline y_2 y_1 )   \leqno (7)     $$
    $$ - ( \overline x_2 x_1 + \overline y_2 y_1 ) 
       ( c^+_{i\uparrow} c_{i\uparrow} +
     c^+_{i\downarrow} c_{i\downarrow} +
     c^+_{j\uparrow} c_{j\uparrow} +
     c^+_{j\downarrow} c_{j\downarrow} )  $$
 $$   - ( \overline x_2 y_1 + \overline y_2 x_1 ) 
   ( c^+_{i\uparrow} c_{j\uparrow} +
     c^+_{i\downarrow} c_{j\downarrow} +
     c^+_{j\uparrow} c_{i\uparrow} +
     c^+_{j\downarrow} c_{i\downarrow} )    $$
we can deduce the formula:
    $$ < X | A_{ij}(x_2,y_2)  A^+_{ij}(x_1,y_1) | Y >   = 
 2 ( \overline x_2 x_1 + \overline y_2 y_1 )  < X | Y >     $$
for any {\it partial} coverings  $ | X > , | Y > $ such that the sites
$ i, j $ are not occupied (i.e. coverings of the lattice {\it without}
the sites i and j ).
 \par
          Analogously, from the commutator  $(k \ne j )$:
   $$ [A_{ik} (x_2,y_2) , A^+_{ij} (x_1,y_1)] =  \overline x_2 x_1
- \overline x_2 x_1   ( c^+_{i\uparrow} c_{i\uparrow} +
     c^+_{i\downarrow} c_{i\downarrow} )
- \overline x_2 y_1   ( c^+_{j\uparrow} c_{i\uparrow} +
     c^+_{j\downarrow} c_{i\downarrow} )  \leqno (9) $$
$$ - \overline y_2 x_1   ( c^+_{i\uparrow} c_{k\uparrow} +
     c^+_{i\downarrow} c_{k\downarrow} )
- \overline y_2 y_1   ( c^+_{j\uparrow} c_{k\uparrow} +
     c^+_{j\downarrow} c_{k\downarrow} )   $$
we obtain:
   $$  < X | A_{jl}(x_3,y_3) A_{ik}(x_2,y_2) A^+_{ij}(x_1,y_1) | Y >
      \ = \ < X | A_{lk}(\overline x_1 x_2 x_3, -\overline y_1 y_2 y_3 ) 
           | Y > \leqno (10) $$
where in the partial
 coverings $ | X >, | Y > $ the sites $ i,j,k,l$ are empty.\par
        In Fig. 1 it is shown how formulas (8),(10), can be used 
to simplify a Sutherland's graph. \par
        From relations (8),(9), it follows that the contribution of a
loop with $m$ nearest-neighbor bonds is given by (we assume $x,y$ real):
     $$  l = 2 x^m + (-1)^{1+ {m\over 2} } 2 y^m  \leqno (11) $$
in agreement with previous results [10],[11]. \par
        Let us now denote by $ A^+_{ij} (\tilde x, \tilde y) $ the 
operator creating a "knight's move bond" ( KM bond) between the sites
$i$ and $j$ whose relative distance is $\sqrt 5$ $(1^2+2^2=5)$!.
 The introduction of
a KM bond simply modifies the parameters $x,y$ into $\tilde x, \tilde y$;
therefore the contribution of a loop containing $r$ KM bonds and 
$m-r$ nearest-neighbor bonds is given by:
     $$  l = 2 x^{m-r} \tilde x^r + (-1)^{1+ {m\over 2} }
             2 y^{m-r} \tilde y^r  \leqno (12) $$
        Formulas (11),(12) allow us to compute the overlap matrix
$ < \phi_i | \phi_j > $. \par
        Let us now compute the matrix elements $ < \phi_r | H \phi_s > $
of the Hubbard Hamiltonian.  Denoting by $ T_{ij}$ the operator:
      $$ T_{ij} =   c^+_{i\uparrow} c_{j\uparrow} +
                    c^+_{i\downarrow} c_{j\downarrow} +
                    c^+_{j\uparrow} c_{i\uparrow} +
                    c^+_{j\downarrow} c_{i\downarrow}    \leqno (13) $$
the Hubbard Hamiltonian can be written as: 
      $$  H = - \sum_{ < ij > } T_{ij} + U \sum_i n_{i \uparrow}
                              n_{i \downarrow}  \leqno (14) $$
where $ \sum_{ < ij > } $ denotes the sum over all nearest-neighbor
pairs. \par
         As for the matrix elements of the operator $T_{ij}$, we must
distinguish three cases: \par
i)  The operator $T_{ij}$ acts on a pair of sites $i,j$ belonging to
two different closed loops.  Since the total particle number must be
conserved inside a loop, the contribution vanishes. The corresponding
graph is shown in Fig.2 .\par
ii) The operator $T_{ij}$ acts on a pair of sites connected by a bond.
For any partial coverings  $ | X >, | Y > $ such that the sites $i,j$
are empty, we have :  
  $$ < X | T_{ij} A^+_{ij} (x,y) | Y > = 2 < X | A^+_{ij} (y,x) | Y >
           \leqno (15) $$
         Hence the hopping operator exchanges the covalent and ionic
bonds.\par
iii)  The operator $T_{ij}$ divides a loop into two parts of lengths
$m,n$; if $m$ and $n$ are odd, by repeated application of Eq. (10)
we reduce ourselves to the preceding case, and Eq.(15) still holds;
if $m$ and $n$ are even we can repeat the same procedure. In this way
we arrive to an expression of the form: 
$$ < 0 | A_{il} (x_1,y_1) A_{jm} (x_2,y_2)  T_{ij}
       A^+_{lj} (x_3,y_3) A^+_{im} (x_4,y_4) | 0 >  \leqno (16) $$
whose contribution vanishes. The corresponding graph is shown
in Fig. 3. \par
     Let us denote by  $V_{ij} =  U  (n_{i \uparrow} n_{i \downarrow} +
   n_{j \uparrow} n_{j \downarrow} )$ the contribution to the potential
energy of the sites $i$ and $j$. Since $V_{ij}$ counts the double
occupations, the action of $V_{ij}$ on $A^+_{ij} (x,y) $ is simply
to annihilate the covalent part of the bond.  Hence:
$$ < X | V_{ij} A^+_{ij} (x,y) | Y > =
    U  < X | A^+_{ij} (x,0) | Y >  \leqno (17) $$
      Let us add a few words on the modifications of the formalism that
are necessary to cover cases different from half-filling. In general,
the scalar product of two coverings can still be easily computed.
However, the formula $ <\phi_i|\phi_j> = \prod_{k=1}^m  l_k $ does not
hold any more. The Sutherland's graph  corresponding to $<\phi_i|\phi_j>$
contains in general
 not only closed loops, but also open paths. These paths can be
easily simplified by using repeatedly formula (10). If the number of 
bonds of a given open path is odd, the final result vanishes, since
$ < A_{i,j} (x,y) > = 0 $ ; if it is even, the final contribution is
of the type : 
    $$ < A_{i,j}(x_1,y_1) A_{j,k}^+ (x_2,y_2) > = \overline {x_1} x_2 . $$
       Notice that only the ionic coefficients $x_1,x_2$, enter in the 
last formula . \par
       The generalization of the rules to be used in order to compute
$< \phi_i | H | \phi_j >$  is  straightforward.
 \par
\spazio 1 
{\bf Numerical results and conclusions.} \par
        In order to obtain a first insight about the values of the
variational parameters,  the scalar product with the exact ground
state, etc., we have first  considered a toy model, i.e. the 
$ 2 \times 2 \times 2 $ cube, which is topologically equivalent
to the $ 4 \times 2 $ lattice  (see e.g. ref. [14] for the case
of the Heisenberg model ).\par
         Denoting the vertices of the cube by the numbers
$ 1,2, \dots ,8 $, we can choose five coverings that give rise to
five different orbits of the space group; in Fig.4 the coverings that
generate these five orbits are shown: there are two coverings without
KM bonds, and three coverings with 1, 2 and 3  KM bonds respectively. \par
   Since we are interested in the totally symmetric state, all the
covering structures  of an orbit are taken with equal weight. \par
     Notice that the KM  correspond to bonds between opposite vertices
of the cube.   The matrix elements  $ S_{rs}, T_{rs}, V_{rs}
(r,s =1,2,..5) $ of the overlap, hopping and interaction matrices
can be easily computed (see Appendix), where the indexes $r,s$
run over the five orbits of Fig.4 . \par
      In Table I we give, for the $ 2 \times 2 \times 2 $ case, the value
of the energies  of the knight's move resonating valence bond   state 
$|\rm{KM}>$ and the nearest neighbor resonating valence bond state $\rm{|NN>}$,
 compared with the exact energy  $ E $  and the best
Hartree-Fock energy  $ \rm{E_{HF}} $; we denote by $\rm{|E>}$ the exact ground
state and we give the values of
 the overlaps  $\rm{ <E|NN>} $, $\rm{ <E|KM>} $.\par
        Even in this toy model we can learn something: the RVB 
variational states  (nearest-neighbor and knight's move) are worse
than the Hartree-Fock state for small $U$, but are superior  
(especially the KMRVB state) for $U > 8 $. \par
         We have repeated the same calculation for the more interesting
case of the  $ 4 \times 4 $ lattice.  For this model there are $272$ \ 
NNRVB coverings; this number increases to $40320$  if we include
also  KMRVB  coverings. In Fig. 5  an exemple of covering of the
 lattice containing 3 KM bonds is shown.\par
We know from previous works
(see, e.g. ref.[13]) that the ground state is totally symmetric with
respect to the space group $G$.  Averaging over $G$, we can reduce
the dimension of the secular problem to 13 in the nearest-neighbor
case, and to $458$ if KM bonds are present. For comparison, the dimension of the
whole Hilbert space is  $ {16\choose 8}^2 \simeq 1.6 \ {10}^8 $,
and the dimension of the totally symmetric subspace
 is $\simeq 1.3 \ {10}^6$ .\par
         In Table II the values of the ground state energies and the
overlaps with the ground state are given for the $ 4 \times 4 $ case.
The notations are the same as those used in Table I.  \par
         From Table II we see that for large U (e.g. for U=40)
 the KMRVB state has an energy value (-1.8973 compared to -1.9084)
and an overlap with the exact ground state of 0.999. Both are
very  satisfactory values. \par
         It is also interesting to remark that considering the projection of the
exact ground state wave function $ | E > $ in the subspace generated by the
coverings $ | \phi_k > $,  the absolute values of the coefficients
$ c_k = < \phi_k | E > $  varies very little  
  ( typical values are 0.025, 0.0248, 0.0253, 0.0269 );
furthermore their sign
 agrees with the Marshall-Peierls
rule; this kind of behaviour is well known for the Heisenberg
model (see ref. [14]). The variational state considered in refs.[10],[11]
consists in taking all the $|c_k|$ equal; this approximation is very
good in the restricted  class of the NNRVB states. \par
          In Table XXXX the variations of the ionicity parameters
$ r= {x\over y} , \tilde r = {\tilde x \over \tilde y} $ with $U$ are exibited
for the models  $ 2 \times 2 \times 2 $ and  $ 4 \times 4 $ . \par
 From the Table we see that the values of the variational parameters
are not exactly the same for a given $U$, but their order of
magnitude is the same. Hence we can speak of an " approximate
transferability " from one model to another. \par
        In table III  we show the results for the spin-spin correlation
function   $ < S_z(0) S_z(\vec r) >$ for $ \vec r = (0,0),(1,0),(2,0),
(2,1),(2,2) $; (in order to save space, in the table only the value of
$|r|=\sqrt{r_x^2+r_y^2}$ is indicated). \par
  As is well known the antiferromagnetic order of the 
Hartree-Fock state is too strong compared with the exact ground state
values, since the best mean field solution is a spin-density wave;
 the antiferromagnetic order of the NNRVB state is  reasonable but
decays too fast as expected  (see also ref  [8] ). The
 spin-spin correlation function derived from the KMRVB state
is almost exact for U=40, but even for U=4 it reproduces the correct
behaviour, while this is not the case for the Hartree-Fock solution.\par
         In table IV we have examined the d-wave pair correlation
function in order to see the effect of the approximation 
on a physical quantity which is probably crucial in high $T_c$
superconductivity. This correlation function is defined as
$ < \Delta^+(0)  \Delta (\vec r) >$  where:  \par
$$  \Delta (i) = \sum_{u=\pm x,\pm y}   f(u)
       ( c_{i \uparrow} c_{i+u \downarrow} +
         c_{i+u \uparrow} c_{i \downarrow} )    , $$
and denoting by $  x $ and  $ y $ unit vectors in the two lattice
directions, $ f(\pm x ) = 1 $  and $ f(\pm y ) = -1 $.
  Both correlation functions, derived from
the NNRVB and KMRVB wave functions, show a quite reasonable
behaviour for the  rate of  decay at increasing values of $|r|$.
We have also computed the s-wave pair correlation function which 
shows a similar behaviour. This kind of agreement for
the s and d-wave pair correlation functions is present only for large
values of U.  No indications of ODLRO appears. \par
        Concluding, the type of variational wave functions  studied
in this work certainly captures a good part of the physics of
the Hubbard model, especially for large values of U. Hence, it would
be of interest to study the excited states, obtained by substituting
one triplet bond for a singlet,
 or to introduce  holes in the system.
A strong limitation of the formalism is constituted by the non
orthogonality of the coverings.
\vfill\eject

\mathsurround=0pt
\newdimen\digitwidth          
\setbox5=\hbox{\rm-}          
\digitwidth=\wd5              
\catcode`?=\active            
\def?{\kern\digitwidth}       %

\newcount\scat
\newcount\bul
\noindent{\bf Appendix}
\bigskip
\def\tx{\tilde x}
\def\ty{\tilde y}
In the following, the matrix elements $S_{rs}, T_{rs}, V_{rs}\quad 
(r,s = 1,2,..5)$ of the overlap,
hopping and interaction matrices for the 2x2x2 cube are given.
 $(\tx,\ty)$ are the parameters of KM bonds,
 $(x,y)$ are the parameters of NN bonds. 
\bigskip\centerline{OVERLAP MATRIX}\nobreak
$$\eqalignno{
S_{11}&={1\over 3}[1+2(x^2-y^2)^2]\cr\cr
S_{12}&={1\over 3}(x^2-y^2)[x^4+y^4+2]\cr\cr
S_{13}&={1\over 3}[6(x^5\tx+y^5\ty)]\cr\cr
S_{14}&={1\over 3}[2(x^2\tx^2-y^2\ty^2)+4(x^6\tx^2-y^6\ty^2)]\cr\cr
S_{15}&=[2(x^2\tx^2-y^2\ty^2)]^2\cr\cr
S_{22}&={1\over 6}[1+(x^2-y^2)^2+(x^6+y^6)]\cr\cr
S_{23}&= {1\over 6}[6(x^3\tx^-y^3\ty]+6(x^7\tx-y^7\ty)]\cr\cr
S_{24}&= {1\over 6}[8(x^4\tx^2+y^4\ty^2)+8(x^3\tx-y^3\ty)^2]\cr\cr
S_{25}&= 2(x^4\tx^4-y^4\ty^4)\cr \cr
S_{33}&={1\over 8}[1+ 6(x^4\tx^2+y^4\ty^2)+
2(x^6+y^6)+12(x^3\tx-y^3\ty)^2]\cr \cr
S_{34}&= (x^3\tx-y^3\ty)+(x^5\tx^3-y^5\ty^3)\cr\cr
S_{35}&= 2(x^3\tx^3+y^3\ty^3)\cr \cr
S_{44}&= {1\over 6}[1+[2(x^2\tx^2-y^2\ty^2)]^2+8(x^4\tx^2+y^4\ty^2)]\cr\cr
S_{45}&= 2(x^2\tx^2-y^2\ty^2)\cr\cr
S_{55}&=1&\cr}$$ 
\bigskip\goodbreak\centerline{HOPPING MATRIX}\nobreak
$$\eqalignno{
T_{11}&={32\over 3}xy(1+2(x^2-y^2)^2)\cr\cr
T_{12}&={16\over 3}xy[2(x^6-y^6)+(xy)^2 (y^2-x^2)+4(x^2-y^2)]\cr\cr
T_{13}&={1\over 3}[16 xy(x^5\tx+y^5\ty)+20 xy(x^3\tx+y^3\ty)-
8x^2y^2(y\tx+x\ty)]\cr\cr
T_{14}&={1\over 3}\{2[24xy(x^4\tx^2-y^4\ty^2)+16\tx\ty x^2y^2
(y^2-x^2)+8x^3y^3(\ty^2-tx^2)]+\cr
&+32xy(x^2\tx^2-y^2\ty^2)
+8xy(\tx^2-\ty^2)\}\cr\cr
T_{15}&=32[xy(\tx^2-\ty^2)(x^2\tx^2-y^2\ty^2)]\cr\cr
T_{22}&= {16\over 3}[1+4(x^4+y^4)-3(xy)^2+2(x^6+y^6)]\cr\cr
T_{23}&={1\over 2}[32xy(x^3\tx-y^3\ty)+12xy(x\tx-y\ty)+ \cr
         &+28xy(x^5\tx-y^5\ty)+12x^2y^2(-x^3\ty+y^3\tx)+
         8x^3y^3(-x\tx+y\ty)]\cr\cr
T_{24}&= {1\over 3}\big[2[16xy(x^4\tx^2-y^4\ty^2)+16xy(x^2\tx^2+y^2\ty^2)-
         16x^2y^2\tx\ty+\cr
         &+xy(x^2\tx^2+y^2\ty^2)]+48xy(x\tx-y\ty)(x^3\tx-y^3\ty)\big]\cr\cr
T_{25}&= 16[xy(x^2\tx^4-y^2\ty^4)+xy\tx^2\ty^2(y^2-x^2)+
         x^2y^2\tx\ty(\ty^2-\tx^2)]\cr\cr
T_{33}&= {1\over 8}\big[24xy+24xy(x^4+y^4)+3[16xy(x^4\tx^2+y^4\ty^2)
         +16xy(x^2\tx^2+y^2\ty^2)+\cr
         &+4x^2y^2\tx\ty]+3\cdot 48 xy(x\tx-y\ty) (x^3\tx-y^3\ty)\big]
\cr\cr
T_{34}&= {1\over 2}[12xy(x\tx-y\ty)+16xy(x^3\tx-y^3\ty)+
         20 xy(x^3\tx^3-y^3\ty^3)+\cr
         &+4xy\tx\ty(-x^3\ty-y^3\tx) +24x^2y^2\tx\ty(-x\tx+y\ty)]\cr\cr
T_{35}&= 12[xy(x\tx^3+-y\ty^3)-x\tx y\ty(x\ty+y\tx)]\cr\cr
T_{44}&= {1\over 6}[16xy+4(16xy(x^2\tx^2+y^2\ty^2)-16x^2y^2\tx\ty)  
         +32xy(x^2\tx^2-y^2\ty^2)(\tx^2-\ty^2)]\cr\cr
T_{45}&= 8xy(\tx^2-\ty^2) \cr\cr
T_{55}&= 0 &\cr}$$
\goodbreak\bigskip\centerline{INTERACTION MATRIX}\nobreak
$$\eqalignno{
V_{11}&= {8\over 3}[1+2x^2(x^2-y^2)]\cr\cr
V_{12}&= {8\over 3}x^4(1+5x^4)\cr\cr
V_{13}&= 4x^2(x^5\tx+y^5\ty)+6x^5\tx\cr\cr
V_{14}&={1\over 3}[16x^6\tx^2+8x^2(x^2\tx^2-y^2\ty^2)+4x^2\tx^2]\cr\cr
V_{15}&=16x^2\tx^2(x^2\tx^2-y^2\ty^2)\cr\cr 
V_{22}&={4\over 3}[1+2(x^6+y^6)+6x^4+x^2(x^2-y^2)]\cr\cr 
V_{23}&= {1\over 2}[8x^2(x^2\tx^2-y^2\ty^2) +4x^2\tx^2+8x^7\tx]\cr\cr
V_{24}&= {1\over 3}[12 x^4\tx^2+8x^2(x^4\tx^2+y^4\ty^2)+
        16x^3\tx(x^3\tx-y^3\ty)]\cr\cr
V_{25}&= 16x^4\tx^4\cr\cr
V_{33}&= {1\over 8}[12x^2(x^4\tx^2+y^4\ty^2)+9x^4\tx^2+
        48x^3\tx(x^3\tx-y^3\ty) +\cr
        &+ 6x^6+4\tx^2(x^6+y^6)+6x^2+2\tx^2]\cr\cr
V_{34}&= 2x^3\tx+2(\tx^2+x^2)(x^3\tx-y^3\ty)+8x^5\tx^3\cr\cr
V_{35}&= 4\tx^2(x^3\tx^3+y^3\ty^3)+6x^3\tx^3\cr\cr
V_{44}&= {1\over 3}[x^2+2\tx^2+12x^4\tx^2+8\tx^2(x^4\tx^2+y^4\ty^2) 
         +8x^2\tx^2(x^2\tx^2-y^2\ty^2)]\cr\cr
V_{45}&= 8\tx^2(x^2\tx^2-y^2\ty^2)+4x^2\tx^2\cr\cr
V_{55}&= 8\tx^2&\cr}$$
\vfill\eject

\par
\spazio 2
\par
{\centerline{\bf References}}
\vskip 0.7 true cm
\item{[1]} R.B.Laughlin,  Phys.Rev.Lett. {\bf 50},1395 (1983);
 Phys. Rev. B {\bf 27}, 3383 (1983).
\vskip 0.2 true cm
\par\noindent             
\item{[2]} J.K.Jain,  Phys.Rev.Lett. {\bf 63},199 (1989);
 Phys. Rev. B {\bf 40}, 8079 (1989); B {\bf 41},7653 (1990).
\vskip 0.2 true cm
\par\noindent             
\item{[3]} M.C. Gutzwiller,  Phys.Rev.Lett. {\bf 10},159 (1963);
 Phys. Rev. A {\bf 137}, 1726 (1965).
\vskip 0.2 true cm
\par\noindent             
\item{[4]} P.W.Anderson,  Science {\bf 235},1196 (1987).
\vskip 0.2 true cm
\par\noindent
\item{[5]} C.Gros, Phys. Rev. B {\bf 38}, 931 (1988);
Annals of Phys. {\bf 189},53 (1989).
\vskip 0.2 true cm
\par\noindent
\item{[6]} B. Sutherland, Phys. Rev. B {\bf 37}, 3786 (1988).
\vskip 0.2 true cm
\par\noindent
\item{[7]} T.Oguchi and H.Kitatani, J.Phys. Soc. Jpn. {\bf 58},
1403 (1988). 
\vskip 0.2 true cm
\par\noindent
\item{[8]} J.H.Choi and W.R.Thorson, J.Chem.Phys. {\bf 57},252 (1971).
\vskip 0.2 true cm
\par\noindent
\item{[9]} A.Messager and J.L.Richard, Phys.Lett. A {\bf 143},345 (1990).
\vskip 0.2 true cm
\par\noindent
\item{[10]} G.M.Cicuta and S.Stramaglia, Phys.Lett. A {\bf 165},456 (1992).
\vskip 0.2 true cm
\par\noindent
\item{[11]} S.Liang, B.Doucot, P.W.Anderson, Phys. Rev. Lett.
{\bf 61},365 (1988).
\vskip 0.2 true cm
\par\noindent
\item{[12]} W.Marshall Proc. Roy. Soc. A {\bf 232},48 (1955).
\vskip 0.2 true cm
\par\noindent
\item{[13]} G.Fano, F.Ortolani and A.Parola, Phys.Rev.B {\bf 46},1048 (1992).
\vskip 0.2 true cm
\par\noindent
\item{[14]} P.L.Iske and W.J.Caspers, Physica {\bf 142 A},360 (1987).
\vfill \eject
Table I.  Energy values of the ground state, Hartree-Fock and RVB states
for the $ 2\times 2 \times 2$ lattice.
\midinsert
\vbox{
  \def\r {\noalign{\hrule}}
  \def\s {\omit&height2pt& && && && && && && &\cr}
  \def\n {&\cr\s\r\s}
  \def\e {&\cr\s\r}
  \def\f {\hfil}
  \def\b {\langle}
  \def\k {\rangle}
  \offinterlineskip
  \halign{\strut#&\vrule#&&\quad\f$#$\quad&\vrule#\cr
%
%
\r\s
&&  U && E\f      &&E_{HF}\f  &&E_{NN}\f &&E_{KM}\f &&\b E|NN\k&&\b E|KM\k\n
&&  0 && -12.0000 && -12.0000 && -8.8000 && -9.3463 && 0.5477  && 0.6941  \n
&&  1 && -10.1187 && -10.0000 && -7.2887 && -7.7171 && 0.5097  && 0.6765  \n
&&  4 &&  -5.9542 &&  -5.0274 && -4.8152 && -5.0079 && 0.6887  && 0.8101  \n
&&  8 &&  -3.4671 &&  -2.8341 && -3.1667 && -3.2697 && 0.9218  && 0.9763  \n
&& 16 &&  -1.8772 &&  -1.4772 && -1.8002 && -1.8464 && 0.9760  && 0.9986  \n
&& 40 &&  -0.7763 &&  -0.5985 && -0.7553 && -0.7728 && 0.9850  && 0.9993  \e
%
}}
\endinsert
\bigskip
Table II.  Energy values of the ground state, Hartree-Fock and RVB states
for the $4 \times 4 $ lattice.

\midinsert
\vbox{
  \def\r {\noalign{\hrule}}
  \def\s {\omit&height2pt& && && && && && && &\cr}
  \def\n {&\cr\s\r\s}
  \def\e {&\cr\s\r}
  \def\f {\hfil}
  \def\b {\langle}
  \def\k {\rangle}
  \offinterlineskip
  \halign{\strut#&\vrule#&&\quad\f$#$\quad&\vrule#\cr
%
%
\r\s
&&  U && E\f      && E_{HF}\f && E_{NN}\f && E_{KM}\f &&\b E|NN\k&&\b E|KM\k\n
&&  1 && -20.7936 && -20.6542 && -15.5266 && -16.9183 && 0.44317 && 0.64978 \n
&&  4 && -13.6218 && -12.5665 && -10.7020 && -11.5642 && 0.58962 && 0.79090 \n
&&  8 &&  -8.4689 &&  -7.3896 &&  -7.2782 &&  -7.7707 && 0.78905 && 0.94436 \n
&& 16 &&  -4.6119 &&  -3.9116 &&  -4.2421 &&  -4.4830 && 0.89713 && 0.99489 \n
&& 40 &&  -1.9084 &&  -1.5941 &&  -1.8051 &&  -1.8973 && 0.92734 && 0.99904 \e
%
}}
\endinsert
\bigskip
\vfill\eject
Table III. Values of the spin-spin correlation function for the exact
ground state, the Hartree-Fock and the RVB states.
\midinsert
\vbox{
  \def\r {\noalign{\hrule}}
  \def\o{\omit}
  \def\rr{\o&\o&\o&\multispan{11}\hrulefill\cr}
  \def\s {\omit&height2pt& && && && && && &\cr}
  \def\n {&\cr\s\r\s}
  \def\nn{&\cr\s\rr\s}
  \def\e {&\cr\s\r}
  \def\f {\hfil}
  \def\b {\langle}
  \def\k {\rangle}
  \def\sf{\sqrt{5}}
  \def\st{\sqrt{2}}
  \offinterlineskip
  \halign{\strut#&\vrule#&&\quad\f$#$\quad&\vrule#\cr
\r
\omit&height2pt&\multispan{11}&\cr
&&              \multispan{11}\f $\b S_z(0)S_z(r)\k$\f&\cr
\omit&height2pt&\multispan{11}&\cr
%
%
\r\s
&&\o   && |r|&&{HF} &&{NNRVB}&& {KMRVB}&&{GROUND}\n
&&\o   &&  0 &&  0.18704 &&  0.21316 &&  0.20883 &&  0.19244 \nn
&&\o   &&  1 && -0.13706 && -0.08782 && -0.08841 && -0.06879 \nn
&&U=4  &&  2 &&  0.12321 &&  0.03401 &&  0.05120 &&  0.03750 \nn
&&\o   &&\sf && -0.12501 && -0.02010 && -0.05170 && -0.04520 \nn
&&\o   &&2\st&&  0.12201 &&  0.01447 &&  0.04442 &&  0.03851 \n
&&\o   &&  0 &&  0.24876 &&  0.24863 &&  0.24855 &&  0.24853 \nn
&&\o   &&  1 && -0.24784 && -0.11094 && -0.11579 && -0.11590 \nn
&&U=40 &&  2 &&  0.24753 &&  0.05014 &&  0.07110 &&  0.07085 \nn
&&\o   &&\sf && -0.24753 && -0.03262 && -0.06811 && -0.06750 \nn
&&\o   &&2\st&&  0.24753 &&  0.02476 &&  0.06048 &&  0.05997 \e
%
}}
\endinsert
Table IV. Values of the pair correlation function for the exact
ground state,  and the RVB states.

\midinsert
\vbox{
  \def\r {\noalign{\hrule}}
  \def\s {\omit&height2pt& && && && &\cr}
  \def\n {&\cr\s\r\s}
  \def\e {&\cr\s\r}
  \def\f {\hfil}
  \def\b {\langle}
  \def\k {\rangle}
  \def\sf{\sqrt{5}}
  \def\st{\sqrt{2}}
  \offinterlineskip
  \halign{\strut#&\vrule#&&\quad\f$#$\quad&\vrule#\cr
\r
\omit&height2pt&\multispan{7}&\cr
&&   \multispan{7}\f  $\b\Delta^\dagger(0)\Delta(r)\k$ $d$-wave, $U=40$\f&\cr
\omit&height2pt&\multispan{7}&\cr
%
%
\r\s
&& |r|&&{NNRVB}&& {KMRVB}&&{GROUND}\n
&&  0 && 2.33389 && 2.39191 && 2.38947 \n
&&  1 && 0.59034 && 0.60339 && 0.60118 \n
&&  2 && 0.00199 && 0.00114 && 0.00252 \n
&&\sf && 0.00002 && 0.00003 && 0.00005 \n
&&2\st&& 0.00000 && 0.00000 && 0.00000 \n
&&\st && 0.00589 && 0.00487 && 0.00261 \e
%
}}
\endinsert

\vfill\eject
\centerline                {     FIGURE CAPTIONS  }
\noindent
Fig.1  The operators  $ A_{ij}$(dotted line)and $A^+_{ij}$ (full line)
are represented by a bond connecting the points i and j. \par
\spazio 1
\noindent
Fig.2  Graph representing the operator $T_{ij}$ acting
 on a pair of sites $i,j$ belonging to
two different closed loops. The corresponding contribution vanishes. \par
\spazio 1
\noindent
Fig.3  Graph corresponding to the expectation value :
$$ < 0 | A_{il} (x_1,y_1) A_{jm} (x_2,y_2)  T_{ij}
       A^+_{lj} (x_3,y_3) A^+_{im} (x_4,y_4) | 0 >   $$
whose contribution vanishes . \par
\spazio 1
\noindent
Fig.4  Representative coverings of the $ 2 \times 2 \times 2 $ lattice
which generate different orbits of the space group. \par
\spazio 1
\noindent
Fig.5  An exemple of covering of the $4 \times 4 $
 lattice containing 3 KM bonds; point 13 and point 16 are
connected by a bond because of periodicity.\par

\vfill\eject

\input pictex.tex
\newcount\bul
\setbox\bul=\hbox{$\bullet$}\par   

\quad\vskip5truecm
\vbox{\hbox to \hsize{\hfil\beginpicture
\setcoordinatesystem units <0.9pt,0.9pt>
\setsolid
\putrectangle corners at -190 -20 and 200 80

\setdashes
\putrule from -151 10 to -151 50
\putrule from 10 50 to 50 50
\putrule from 10 10 to 50 10
\put{\copy\bul} at -150 10
\put{\copy\bul} at -150 50
\put{\copy\bul} at 10 50 
\put{\copy\bul} at 50 50 
\put{\copy\bul} at 10 10
\put{\copy\bul} at 50 10
\setsolid
\putrule from -149 10 to -149 50
\put 1 at -156 30
\put 2 at -142 30
\put{$ = 2 ( \overline x_1 x_2 + \overline y_1 y_2)$} at -90 30 
\putrule from 50 50 to 50 10
\put{\copy\bul} at 50 50
\put{\copy\bul} at 50 10
\put i at 56 56
\put j at 56 4
\put k at 4 56
\put l at 4 4
\setdashes
\putrule from 100 50 to 100 10
\put{\copy\bul} at 100 50
\put{\copy\bul} at 100 10
\put k at 106 56
\put l at 106 4
\put 1 at 57 30 
\put = at 80 30
\put 2 at 30 56
\put 3 at 30 4
\put{$(\overline x_1 x_2 x_3, - \overline y_1 y_2 y_3)$} at 150 30
\endpicture\hfil}}

\vskip.8truecm\centerline{Fig. 1}
\vfill\eject

\quad\vskip5truecm
\vbox{\hbox to \hsize{\hfil\beginpicture
\setcoordinatesystem units <1.3pt,1.3pt> 
\putrectangle corners at -140 -40 and 140 40
\put = at 60 0
\put 0 at 100 0
\setdashes
\circulararc 180 degrees from -20 0  center at 0 0
\circulararc 180 degrees from -100 0 center at -80 0
\setsolid
\circulararc 180 degrees from -60 0  center at -80 0
\circulararc 180 degrees from  20 0 center at 0 0

\putrule from -60 0 to -20 0

\put {$T_{ij}$} at -40 7
\put{\copy\bul} at -60 0
\put{\copy\bul} at -20 0
\put i at -56 -7
\put j at -24 -7
\endpicture\hfil}}
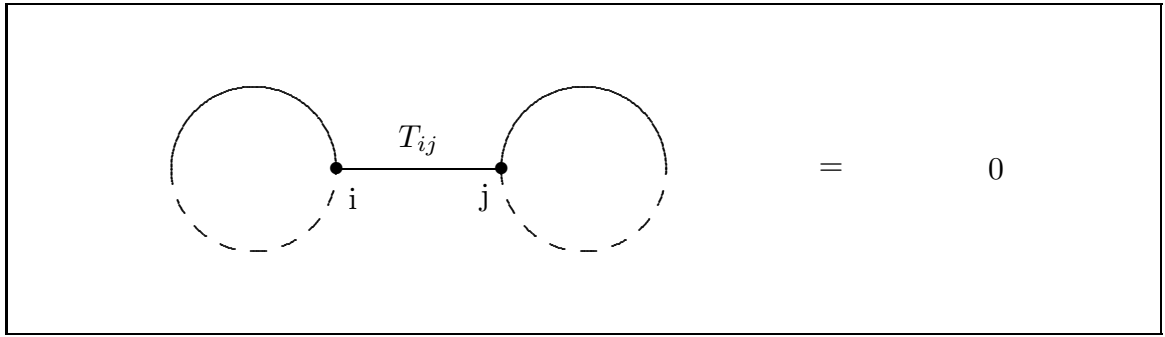
\vskip.8truecm\centerline{Fig. 2}

\vfill\eject

\quad\vskip5truecm
\vbox{\hbox to \hsize{\hfil\beginpicture
\setcoordinatesystem units <1.8pt,1.8pt>
\putrectangle corners at -120 -40 and 80 40
\put = at 0 0
\put 0 at 40 0
\setsolid
\putrule from -80 20 to -40 20
\putrule from -80 -20 to -40 -20
\plot -80 -20 -40 20 /
\setdashes
\putrule from -80 20 to -80 -20
\putrule from -40 20 to -40 -20
\put {$T_{ij}$} at -65 5
\put{\copy\bul} at -40 20
\put{\copy\bul} at -40 -20
\put{\copy\bul} at -80 20
\put{\copy\bul} at -80 -20
\put l at -87 27
\put j at -33 27
\put i at -87 -27
\put m at -33 -27
\put{$1$} at -87 0
\put{$2$} at -33 0
\put{$3$} at -60 27
\put{$4$} at -60 -27

\endpicture\hfil}}
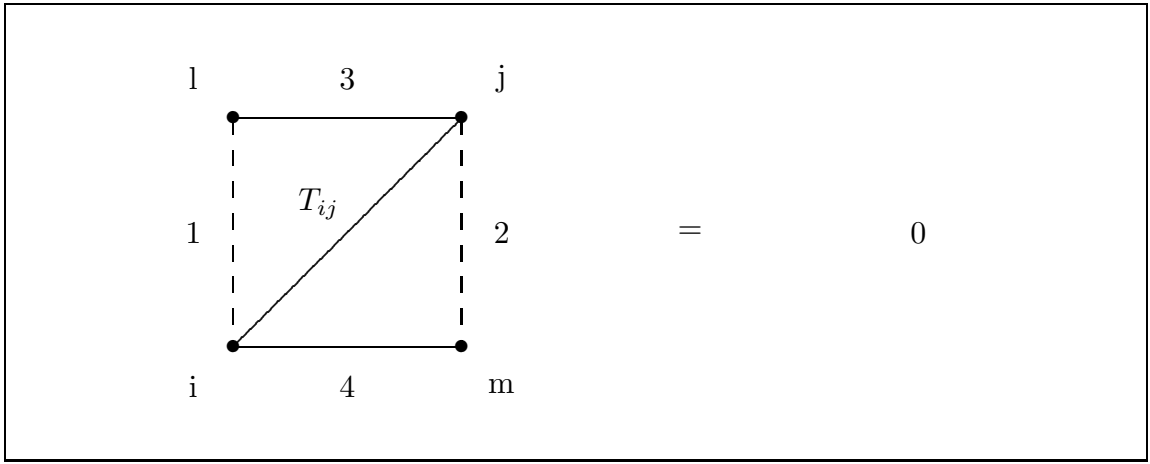
\vskip.8truecm\centerline{Fig. 3}
\vskip1truecm
\vfill\eject

\quad\vskip5truecm
\vbox{
\hbox to \hsize{\hfill
\beginpicture
\setcoordinatesystem units <1.2pt,1.2pt>
\putrectangle corners at -155 50 and 155 -120
\put{
\setplotarea x from 0 to 90 , y from 0 to 80
\setdashpattern <.5pt,2pt>   
\putrectangle corners at 20 20 and 60 60 
\putrectangle corners at 30 30 and 70 70 
\plot 20 20 30 30 /
\plot 60 20 70 30 /
\plot 20 60 30 70 /
\plot 60 60 70 70 /
\put 1 at  16 64
\put 2 at  64 58
\put 3 at       74 74
\put 4 at       26 74
\put 5 at  16 16
\put 6 at  64 16
\put 7 at       74 26
\put 8 at       26 32
\setsolid
\linethickness=.8pt
\putrule from 20 20 to 20 60
\putrule from 30 30 to 30 70
\putrule from 60 20 to 60 60
\putrule from 70 30 to 70 70
\put {\copy\bul} at 20 20
\put {\copy\bul} at 20 60
\put {\copy\bul} at 30 30
\put {\copy\bul} at 30 70
\put {\copy\bul} at 60 20
\put {\copy\bul} at 60 60
\put {\copy\bul} at 70 30
\put {\copy\bul} at 70 70
} at -120 0
\put{
\setplotarea x from 0 to 90 , y from 0 to 80
\setdashpattern <.5pt,2pt>   
\putrectangle corners at 20 20 and 60 60 
\putrectangle corners at 30 30 and 70 70 
\plot 20 20 30 30 /
\plot 60 20 70 30 /
\plot 20 60 30 70 /
\plot 60 60 70 70 /
\put 1 at  16 64
\put 2 at  64 58
\put 3 at       74 74
\put 4 at       26 74
\put 5 at  16 16
\put 6 at  64 16
\put 7 at       74 26
\put 8 at       26 32
\setsolid
\linethickness=.8pt
\putrule from 20 20 to 60 20
\putrule from 30 30 to 30 70
\putrule from 20 60 to 60 60
\putrule from 70 30 to 70 70
\put {\copy\bul} at 20 20
\put {\copy\bul} at 20 60
\put {\copy\bul} at 30 30
\put {\copy\bul} at 30 70
\put {\copy\bul} at 60 20
\put {\copy\bul} at 60 60
\put {\copy\bul} at 70 30
\put {\copy\bul} at 70 70
} at 40 0
\put {
\setplotarea x from 0 to 90 , y from 0 to 80
\setdashpattern <.5pt,2pt>   
\putrectangle corners at 20 20 and 60 60 
\putrectangle corners at 30 30 and 70 70 
\plot 20 20 30 30 /
\plot 60 20 70 30 /
\plot 20 60 30 70 /
\plot 60 60 70 70 /
\put 1 at  16 64
\put 2 at  64 58
\put 3 at       74 74
\put 4 at       26 74
\put 5 at  16 16
\put 6 at  64 16
\put 7 at       74 26
\put 8 at       26 32
\setsolid
\linethickness=.8pt
\plot 30 70  60 20 /
\putrule from 20 20 to 20 60
\plot 60 60  70 70 /
\putrule from 30 30 to 70 30
\put {\copy\bul} at 20 20
\put {\copy\bul} at 20 60
\put {\copy\bul} at 30 30
\put {\copy\bul} at 30 70
\put {\copy\bul} at 60 20
\put {\copy\bul} at 60 60
\put {\copy\bul} at 70 30
\put {\copy\bul} at 70 70
} at -140 -80
\put {
\setplotarea x from 0 to 90 , y from 0 to 80
\setdashpattern <.5pt,2pt>   
\putrectangle corners at 20 20 and 60 60 
\putrectangle corners at 30 30 and 70 70 
\plot 20 20 30 30 /
\plot 60 20 70 30 /
\plot 20 60 30 70 /
\plot 60 60 70 70 /
\put {\copy\bul} at 20 20
\put {\copy\bul} at 20 60
\put {\copy\bul} at 30 30
\put {\copy\bul} at 30 70
\put {\copy\bul} at 60 20
\put {\copy\bul} at 60 60
\put {\copy\bul} at 70 30
\put {\copy\bul} at 70 70
\put 1 at  16 64
\put 2 at  64 58
\put 3 at       74 74
\put 4 at       26 74
\put 5 at  16 16
\put 6 at  64 16
\put 7 at       74 26
\put 8 at       26 32
\setsolid
\linethickness=.8pt
\putrule from 20 20 to 20 60
\plot 60 20 30 70 /
\putrule from 70 30 to 70 70
\plot 30 30  60 60 /
} at -40 -80
\put{
\setplotarea x from 0 to 90 , y from 0 to 80
\setdashpattern <.5pt,2pt>   
\putrectangle corners at 20 20 and 60 60 
\putrectangle corners at 30 30 and 70 70 
\plot 20 20 30 30 /
\plot 60 20 70 30 /
\plot 20 60 30 70 /
\plot 60 60 70 70 /
\put {\copy\bul} at 20 20
\put {\copy\bul} at 20 60
\put {\copy\bul} at 30 30
\put {\copy\bul} at 30 70
\put {\copy\bul} at 60 20
\put {\copy\bul} at 60 60
\put {\copy\bul} at 70 30
\put {\copy\bul} at 70 70
\put 1 at  16 64
\put 2 at  64 58
\put 3 at       74 74
\put 4 at       26 74
\put 5 at  16 16
\put 6 at  64 16
\put 7 at       74 26
\put 8 at       26 32
\setsolid
\linethickness=.8pt
\plot 20 20 70 70 /
\plot 60 20 30 70 /
\plot 30 30 60 60 /
\plot 20 60 70 30 /
} at 60 -80
\endpicture\hfill}
}
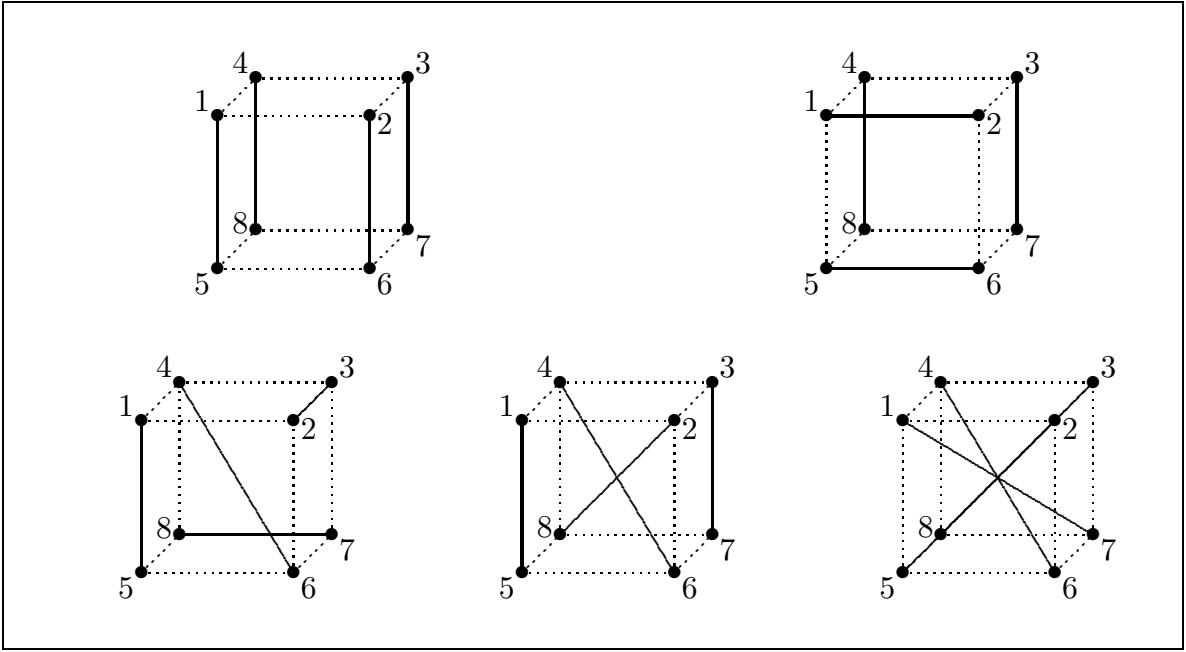
\vskip.8truecm\centerline{Fig. 4}
\vfill\eject

\quad\vskip1truecm
\vbox{\hbox to \hsize{\hfil\beginpicture
\setcoordinatesystem units <6pt,6pt>
\putrectangle corners at -10 -10 and 40 40
\setdashpattern <0.8pt,3.2pt>
\putrule from 0  0 to 0  30
\putrule from 10 0 to 10 30
\putrule from 20 0 to 20 30
\putrule from 30 0 to 30 30
\putrule from 0 0   to  30 0
\putrule from 0 10  to  30 10
\putrule from 0 20  to  30 20
\putrule from 0 30  to  30 30
\put 1 at -1 31
\put 2 at 9 31
\put 3 at 19 31
\put 4 at 29 31
\put 5 at -1 21
\put 6 at 9 21
\put 7 at 21 21
\put 8 at 29 21
\put 9 at -1 11
\put {10} at 8.5 11
\put {11} at 21.5 11
\put {12} at 28.5 11
\put {13} at -1.5 1
\put {14} at 8.5 1
\put {15} at 18.5 1
\put {16} at 28.5 1 

\setsolid
\putrule from 0 -6 to 30 -6
\putrule from 10 0 to 20 0
\putrule from 0 10 to 10 10

\putrule from 10 30 to 20 30
\putrule from 30 10 to 30 20
\put{\copy\bul} at 0 0
\put{\copy\bul} at 0 10
\put{\copy\bul} at 0 20
\put{\copy\bul} at 0 30
\put{\copy\bul} at 10 0 
\put{\copy\bul} at 10 10
\put{\copy\bul} at 10 20 
\put{\copy\bul} at 10 30 
\put{\copy\bul} at 20 0
\put{\copy\bul} at 20 10
\put{\copy\bul} at 20 20
\put{\copy\bul} at 20 30
\put{\copy\bul} at 30 0
\put{\copy\bul} at 30 10
\put{\copy\bul} at 30 20 
\put{\copy\bul} at 30 30
\circulararc 180 degrees from 30 -6 center at 30 -3 

\circulararc 180 degrees from 0 0 center at 0 -3 
\plot 0 30  20 20 / 
\plot 10 20 30 30 /
\plot 0 20 20 10 /

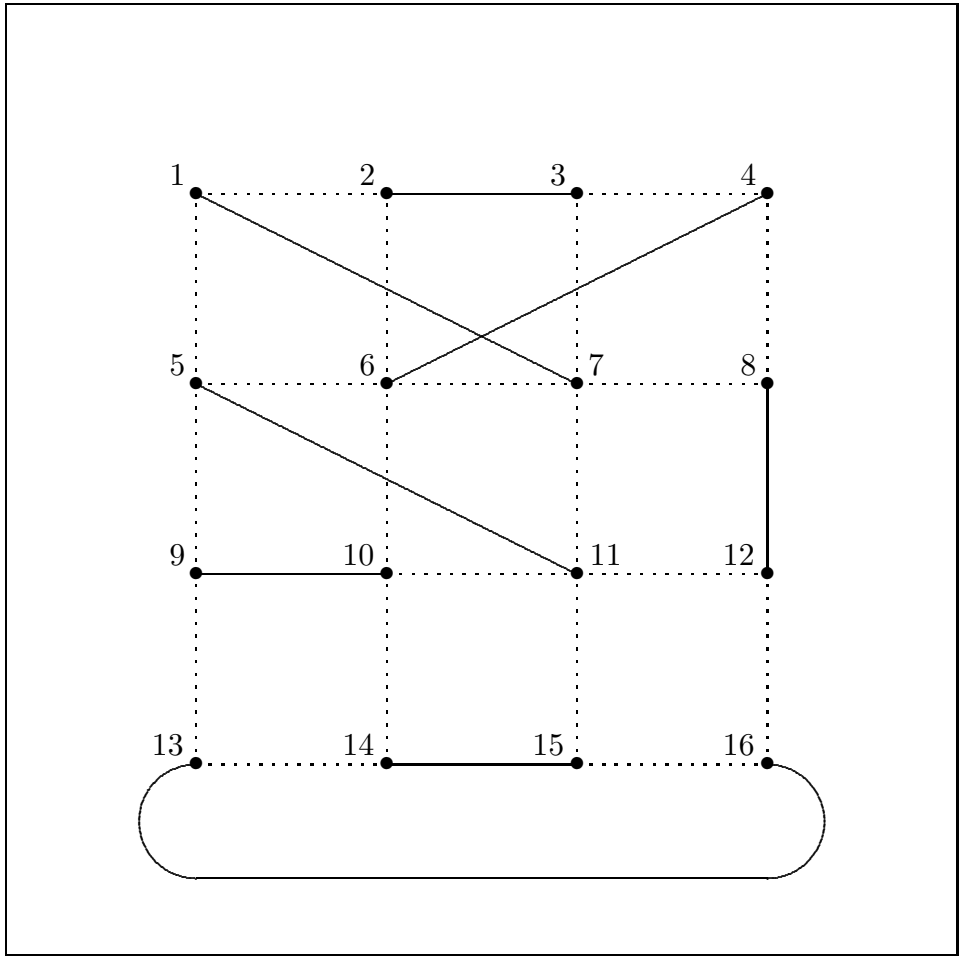
\endpicture\hfil}} \bigskip \centerline{Fig. 5}

\vfill\eject\bye